\documentstyle[preprint,aps,prl,multicol,epsfig]{revtex}

\begin{document}

\newcommand{\beq}{\begin{eqnarray}}
\newcommand{\eeq}{\end{eqnarray}}
\renewcommand{\thefootnote}{\fnsymbol{footnote}}

\title{Nonextensive statistical mechanics - Applications to nuclear and high energy physics \thanks{To appear in the Proceedings of the {\it  X International Workshop on Multiparticle Production - Correlations and Fluctuations in QCD} (8-15 June 2002, Crete), ed. N. Antoniou (World Scientific, Singapore, 2003)}}

\author{
Constantino Tsallis  \\
}
\address{
Centro Brasileiro de Pesquisas F\'\i sicas \\
Rua  Xavier Sigaud 150, 
22290-180 Rio de Janeiro, RJ, Brazil 
}

\author{
Ernesto P. Borges  \\
}
\address{Escola Politecnica,
Universidade Federal da Bahia,
Rua Aristides Novis 2 \\
40210-630 Salvador-BA, Brazil \thanks{tsallis@cbpf.br, ernesto@ufba.br} 
}

\date{\today}
\maketitle
\begin{abstract}

A variety of phenomena in nuclear and high energy physics seemingly do not satisfy the basic hypothesis for possible stationary states to be of the type covered by  Boltzmann-Gibbs (BG) statistical mechanics. More specifically, the system appears to relax, along time, on macroscopic states which violate the ergodic assumption. 
Some of these phenomena appear to follow, instead, the prescriptions of nonextensive statistical mechanics. In the same manner that the BG formalism is based on the entropy $S_{BG}=-k \sum_i p_i \ln p_i$, the nonextensive one is based on the form $S_q=k(1-\sum_ip_i^q)/(q-1)$ (with $S_1=S_{BG}$). Typically, the systems following the rules derived from the former exhibit an {\it exponential} relaxation with time toward a stationary state characterized by an {\it exponential} dependence on the energy ({\it thermal equilibrium}), whereas those following the rules derived from the latter are characterized by (asymptotic) {\it power-laws} (both the typical time dependences, and the energy distribution at the stationary state).
A brief review of this theory is given here, as well as of some of its applications, such as electron-positron annihilation producing hadronic jets, collisions involving heavy nuclei, the solar neutrino problem, anomalous diffusion of a quark in a quark-gluon plasma, and flux of cosmic rays on Earth. In addition to these points, very recent developments generalizing nonextensive statistical mechanics itself are mentioned.

\end{abstract}



\section{introduction}

The foundation of statistical mechanics comes from mechanics (classical, quantum, relativistic, or any other elementary dynamical theory). Consistently, in our opinion, the expression of entropy to be adopted for formulating statistical mechanics (and ultimately thermodynamics) depends on the particular type of occupancy of phase space (or Hilbert space, or Fock space, or analogous space) that the microscopic dynamics of the system under study collectively favors. In other words, it appears to be nowadays strong evidence that statistical mechanics is larger than Boltzmann-Gibbs (BG) statistical mechanics, that the concept of physical entropy can be larger than 
\beq
S_{BG}=-k \sum_i^W p_i \ln p_i 
\eeq
(hence $S_{BG}=k \ln W$ for equal probabilities), pretty much as geometry is known today to be larger than Euclidean geometry, since Euclid's celebrated parallel postulate can be properly generalized in mathematically and physically very interesting manners.

Let us remind the words of A. Einstein \cite{einstein} expressing his understanding of Eq. (1):

\noindent
{\it Usually $W$ is put equal to the number of complexions [...]. In order to
calculate $W$, one needs a {\it complete} (molecular-mechanical) theory of the
system under consideration. Therefore it is dubious whether the Boltzmann
principle has any meaning without a complete molecular-mechanical theory or
some other theory which describes the elementary processes.
$S=\frac{R}{\cal N}\log W+\;{\rm const}.$ seems
without content, from a phenomenological point of view, without giving in
addition such an {\it Elementartheorie}.
} 

This standpoint is, in our opinion, quite similar to the position that Riemann adopted when he began his study of ``the concepts which lie at the base of Geometry". Along this line, it is our understanding that the entropy expressed in Eq. (1)  is, on physical grounds, no more irreducibly universal than Euclidean geometry with regard to all possible geometries, which, by the way, also include the fractal one, which inspired the theory addressed in this paper. Nonextensive statistical mechanics \cite{tsallis1}, to which this brief review is dedicated, is based on the following expression
\beq
S_q=k\frac{1-\sum_i^Wp_i^q}{q-1} \;\;\;(q \in {\cal R}; \;S_1=S_{BG}) \;.
\eeq
It is well known that for microscopic dynamics which relax on an {\it ergodic} occupation of phase space, the adequate entropic form to be used is that of Eq. (1). Such assumption is ubiquitously satisfied, and constitutes the physical basis for the great success of BG thermostatistics since over a century. We believe that a variety of more complex occupations of phase space may be handled with more complex entropies. In particular, it seems that Eq. (2), associated with an index $q$ which is dictated by the microscopic dynamics (and which generically differs from unity), is adequate for a vast class of stationary states ubiquitously found in Nature.
In recent papers, E.G.D. Cohen \cite{cohen} and M. Baranger \cite{baranger} also have addressed this question.  

A significant amount of systems, e.g., turbulent fluids (\cite{turbulentbeck,turbulentarimitsu} and references therein), electron-positron annihilation \cite{bediaga,beckannihilation}, collisions of heavy nuclei \cite{wilk,kodama,ion}, solar neutrinos \cite{superkamiokande,neutrino}, quark-gluon plasma \cite{rafelski}, cosmic rays \cite{cosmic},  self-gravitating systems \cite{polytrope}, peculiar velocities of galaxy clusters \cite{peculiar}, cosmology \cite{cosmology}, chemical reactions \cite{astero}, economics \cite{economics1,economics2,economics3}, motion of {\it Hydra viridissima} \cite{hydra}, theory of anomalous kinetics \cite{ademir}, classical chaos \cite{chaos1}, quantum chaos \cite{chaos2}, quantum entanglement \cite{entanglement}, anomalous diffusion \cite{nonlinearFP}, long-range-interacting many-body classical Hamiltonian systems (\cite{longrange} and references therein), internet dynamics \cite{internet}, and others, are known nowadays which in no trivial way accommodate within BG statistical mechanical concepts. Systems like these have been handled with the functions and concepts which naturally emerge within nonextensive statistical mechanics \cite{tsallis1,tsallis2,tsallis3}. 

We may think of $q$ as a biasing parameter: $q<1$ privileges rare events, while $q>1$ privileges common events. Indeed, $p<1$ raised to a power $q<1$ yields a value {\it larger} than $p$, and the relative increase $p^q/p=p^{q-1}$ is a {\it decreasing} function of $p$, i.e., values of $p$ closer to 0 (rare events) are benefited. Correspondingly, for $q>1$, values of $p$ closer to 1 (common events) are privileged. Therefore, the BG theory (i.e., $q=1$) is the unbiased statistics. A concrete consequence of this is that the BG formalism yields {\it exponential} equilibrium distributions (and time behavior of typical relaxation functions), whereas nonextensive statistics yields (asymptotic) {\it power-law} distributions (and relaxation functions). Since the BG exponential is recovered as a limiting case, we are talking of a {\it generalization}, not an alternative.

To obtain the probability distribution associated with the relevant stationary state (thermal equilibrium or metaequilibrium) of our system we must optimize the entropic form (2) under the following constraints \cite{tsallis1,tsallis2}:
\beq
\sum_ip_i=1 \;,
\eeq
and
\beq
\frac{\sum_i p_i^q E_i}{\sum_ip_i^q} =U_q\;,
\eeq
where $\{E_i\}$ is the set of eigenvalues of the Hamiltonian (with specific boundary conditions), and $U_q$ is a fixed and {\it finite} number.
This optimization yields
\beq
p_i=\frac{[1-(1-q)\beta_q (E_i-U_q)]^{1/(1-q)}}{Z_q} \;,
\eeq 
where
\beq
Z_q \equiv \sum_j [1-(1-q)\beta_q (E_j-U_q)]^{1/(1-q)} \;,
\eeq

and
\beq
\beta_q \equiv \frac{\beta}{\sum_j p_j^q} \;,
\eeq
$\beta$ being the optimization Lagrange parameter associated with the generalized internal energy $U_q$. Equation (5) can be rewritten as
\beq
p_i \propto [1-(1-q)\beta^\prime E_i]^{1/(1-q)} \equiv e_q^{-\beta^\prime E_i} \;,
\eeq
where $\beta^\prime$ is a renormalized inverse ``temperature", and the {\it $q$-exponential function} is defined as $e_q^x \equiv [1+(1-q) x]^{1/(1-q)}=1/[1-(q-1) x]^{1/(q-1)}$ (with $e_1^x=e^x$).  This function replaces, in a vast number of relations and phenomena, the usual BG factor. In particular, the ubiquitous Gaussian distribution $\propto e^{-ax^2}$ becomes generalized into the distribution $\propto e_q^{-a_q x^2} = 1/[1+(q-1) a_q x^2]^{1/(q-1)}$ (fat-tailed if $q>1$). 

\section{Generalizing nonextensive statistical mechanics}

Nonextensive statistical mechanics generalizes the BG theory. It presumably addresses (multi)fractal-like occupancy of phase space at the associated stationary states (e.g., metaequilibrium), instead of the usual, homogeneous, occupancy which satisfies ergodicity. Is there any fundamental reason for stopping here? We do not think so. In fact, roads pointing towards generalizations of (or alternatives for) nonextensive statistical mechanics are already open in the literature. Let us mention here two of them (already exhibiting some successes), namely (i) crossovers between $q$-statistics and $q^\prime$-statistics (\cite{cosmic} and references therein), and (ii) the recently introduced Beck-Cohen superstatistics \cite{beckcohen}. Both of them address the energy distributions corresponding to the stationary states, and are perfectly compatible, as we shall show. More precisely, the first type can be thought as a particular case of the second type. However, statistical mechanics is much more than a distribution correctly corresponding to the stationary state. Indeed, if any given entropy $S(\{p_i\})$ is optimized by a specific distribution $p_i$, {\it all} entropic forms which are monotonic functions of $S$ will be optimized by the {\it same} distribution. Nevertheless, only a very restricted class of such entropic forms can be considered as serious candidates for constructing a full thermostatistical theory, eventually connected with thermodynamics. In particular, one expects the correct entropy to be {\it concave} and {\it stable}. Such is the case \cite{abestable} of $S_q$  as well as of the generalized entropy recently proposed \cite{andreconstantino,stablesuperstatistics} for the just mentioned superstatistics \cite{beckcohen}. We briefly  address these questions in this Section.

Let us first consider the following differential equation:
\beq
\frac{dy}{dx}=ay \;\;\;(y(0)=1) \;.
\eeq
The solution is given by
\beq
y=e^{ax} \;.
\eeq
We can use this result in at least three manners which are of interest in statistical mechanics:

(i) We may refer to the {\it sensitivity to the initial conditions}, and consider $x \equiv t$, where $t$ is time, $y \equiv \xi \equiv \lim_{\Delta x(0) \to 0} \Delta x(t)/\Delta x(0)$, where $\Delta x(t)$ denotes the discrepancy of two initial conditions in a one-dimensional map (or, for higher-dimensional systems, the analogous situation for the axis along which the maximum dynamical instability occurs), and $a \equiv \lambda_1 \ne 0$, where $\lambda_1$ is the Lyapunov exponent. In this case Eq. (10) reads in the familiar form:
\beq
\xi(t)=e^{\lambda_1 \,t}\;.
\eeq

(ii) We may refer to the {\it relaxation} towards the stationary state (thermal equilibrium), and  consider $x \equiv t$, $y \equiv [{\cal O}(t)-{\cal O(\infty)}]/[{\cal O}(0)-{\cal O(\infty)}]$, where ${\cal O}$ is the average value of a generic observable, and $a \equiv -1/\tau <0$, where $\tau$ is a relaxation time. In this case Eq. (10) reads in the typical form:
\beq
\frac{{\cal O}(t)-{\cal O(\infty)}}{{\cal O}(0)-{\cal O(\infty)}}=e^{-t/\tau}\;.
\eeq 

(iii) We may refer to the {\it energy distribution} at thermal equilibrium of a Hamiltonian system, and consider $x \equiv E_i$, where $E_i$ is the energy of the $i$-th microscopic state, $y=Z\,p(E_i)$, where $p$ is the energy probability and $Z$ the partition function, and $-a \equiv \beta$ is the inverse temperature. In this case Eq. (10) reads in the familiar BG form:
\beq
p(E_i)=\frac{e^{-\beta E_i}}{Z} \;\;\;(Z\equiv \sum_je^{-\beta E_j}) \;.
\eeq
This distribution is of course the one that optimizes the entropy $S_{BG}$ under the standard constraints for the canonical ensemble. 

Let us next generalize Eq. (9) as follows:
\beq
\frac{dy}{dx}=ay^q \;\;\;(y(0)=1) \;.
\eeq
The solution is given by
\beq
y=e_q^{ax} \equiv [1+(1-q)ax]^{1/(1-q)} \;,
\eeq
$e_q^x$ being from now on referred to as the $q$-exponential function ($e_1^x=e^x$). The three above possible physical interpretations of such solution now become

(i) For the sensitivity to the initial conditions,
\beq
\xi(t)=e_q^{\lambda_q \,t}= [1+(1-q)\lambda_q\,t]^{1/(1-q)}\;,
\eeq
where $\lambda_q \ne0$ is the generalized Lyapunov coefficient (see \cite{chaos1}), and, at the edge of chaos, $\lambda_q>0$ and $q<1$. 

(ii) For the relaxation,
\beq
\frac{{\cal O}(t)-{\cal O(\infty)}}{{\cal O}(0)-{\cal O(\infty)}}=e_q^{-t_q/\tau}= \frac{1}{[1+(q-1)\tau_q\,t]^{1/(q-1)}}\;,
\eeq 
where $\tau_q>0$ is a generalized relaxation time, and typically $q \ge 1$ \cite{bemski}. 

(iii) For the energy distribution, we get the form which emerges in nonextensive statistical mechanics, namely \cite{tsallis1,tsallis2}
\beq
p(E_i)=\frac{e_q^{-\beta_q^\prime E_i}}{Z_q^\prime}=\frac{(Z_q^\prime)^{-1}}{   [1+(q-1)\beta_q^\prime E_i]^{1/(q-1)}    } \;\;\;(Z_q^\prime\equiv \sum_je_q^{-\beta_q^\prime E_j}) \;,
\eeq
where usually, but not necessarily, $\beta_q^\prime>0$ and $q \ge 1$. 
This distribution is the one that optimizes the entropy $S_q$ under appropriate constraints for the canonical ensemble. 

Let us next unify Eqs. (9) and (14) as follows:
\beq
\frac{dy}{dx}=a_1y+a_qy^q \;\;\;(y(0)=1) \;.
\eeq
This is a simple particular case of Bernoulli equation, and its solution is given by
\beq
y=\Bigl[ \Bigl(    e^{(1-q)a_1x }-1   \Bigr)\frac{a_q}{a_1} +1               \Bigr]^{\frac{1}{1-q}}
\eeq
This solution reproduces Eq. (10) if $a_q=0$, and Eq. (15) if $a_1=0$. It corresponds to a crossover from a $q \ne 1$ behavior at small values of $x$, to a $q=1$ behavior at large values of $x$. The crossover occurs at $x_c\simeq 1/[(q-1)a_1]$ \cite{bemski}.

\section{applications}

Let us now briefly review five recent applications of the ideas associated with nonextensive statistical mechanics to phenomena in nuclear and high energy physics, namely electron-positron annihilation \cite{bediaga,beckannihilation}, collisions of heavy nuclei \cite{wilk,kodama,ion}, the solar neutrino deficit \cite{superkamiokande,neutrino}, quark anomalous diffusion \cite{rafelski}, and the flux of cosmic rays \cite{cosmic}.

\noindent
{\it Electron-positron annihilation:} 

In high energy collisions of an electron with a positron, annihilation occurs and, immediately after, typically two or three hadronic jets are produced. The probability distribution of their transverse momenta is non-Boltzmannian, more strongly so with increasing energy of collision. This phenomenon has defied theoreticians since several decades, particularly since Hagedorn \cite{hagedorn} quantitatively analyzed such data in the frame of BG statistical mechanics. A phenomenological theory has been recently proposed by Bediaga et al \cite{bediaga}, which beautifully fits the data. The fitting parameters are two, namely the temperature $T$ and the entropic index $q$. For each energy collision $E_c$ a set of $(T,q)$ is determined. It numerically comes out that $q$ depends on the energy (like $q(\infty)-q(E_c) \propto E_c^{-1/2}$ for increasingly large $E_c$, and $q(E_c \simeq 0) \simeq 1$), {\it but $T$ does not!} This invariance of $T$ with respect to the collision energy constitutes the central hypothesis of the physical scenario advanced long ago by Hagedorn. This scenario is now confirmed. The ingredients for a microscopic model within this approach have also been proposed \cite{beckannihilation}.

\noindent
{\it Heavy nuclei collisions:}

A variety of high-energy collisions have been discussed in terms of the present nonextensive formalism. Examples include
proton-proton, central Pb-Pb and other  nuclear collisions \cite{wilk,kodama}.  Along related lines,  entropic inequalities applied to pion-nucleon experimental phase shifts have provided strong evidence of nonextensive quantum statistics \cite{ion}. 

\noindent
{\it Solar neutrino problem:} 

The solar plasma is believed to produce large amounts of neutrinos through a variety of mechanisms (e.g., the proton-proton chain). The calculation done using the  so called Solar Standard Model (SSM) results in a neutrino flux over the Earth, which is roughly the {\it double} of what is measured. This is sometimes referred to as the {\it neutrino problem} or the {\it neutrino enigma}. There is by no means proof that this neutrino flux defect is due to a single cause. It has recently been verified that neutrino oscillations do seem to exist  (\cite{superkamiokande} and references therein), which would account for part of the deficit. But it is not at all clear that it would account for the entire discrepancy. Quarati and collaborators \cite{neutrino} argue that part of it -- even, perhaps, an appreciable part of it -- could be due to the fact that BG statistical mechanics is used within the SSM. The solar plasma involves turbulence, long-range interactions, possibly long-range memory processes, all of them phenomena that could easily defy the applicability of the BG formalism. Then they show \cite{neutrino} in great detail how the modification of the ``tail" of the energy distribution could considerably modify the neutrino flux to be expected. Consequently, small departures from $q=1$ (e.g., $|q-1|$ of the order of $0.1$) would be enough to produce as much as $50\%$ difference in the flux. This is due to the fact that most of the neutrino flux occurs at what is called the {\it Gamow peak}. This peak occurs at energies quite above the temperature, i.e., at energies in the tail of the distribution. 

\noindent
{\it Quark diffusion:} 

The anomalous diffusion of a charm quark in a quark-gluon plasma has been analyzed by Walton and Rafelski \cite{rafelski} through both nonextensive statistical mechanical arguments and quantum chromodynamics. The results coincide, as they should, only for $q=1.114$. 

\noindent
{\it Cosmic rays:} 

The flux $\Phi$ of cosmic rays arriving on Earth is a quantity whose measured range is among the widest experimentally known ($33$ decades in fact). This distribution refers to a range of energies $E$ which also is impressive ($13$ decades). This distribution is very far from exponential: See Figs. 1 and 2. This basically indicates that no BG thermal equilibrium is achieved, but some other (either stationary, or relatively slow varying) state, characterized in fact by a power law. If the distribution is analyzed with more detail, one verifies that two, and not one, power-law regimes are involved, separated by what is called the ``knee" (slightly below $10^{16}\;eV$). At very high energies, the power-law seems to be interrupted by what is called the ``ankle" (close to $10^{19}\;eV$) and perhaps a cutoff. The momenta $M_l \equiv \langle (E-\langle E\rangle)^l \rangle= [\int_0^{E_{cutoff}} dE \,\Phi(E)  (E-\langle E\rangle)^l] / [\int_0^{E_{cutoff}} dE \,\Phi(E)]$ ($l=1,2,3$) as functions of the cutoff energy $E_{cutoff}$ (assumed to be abrupt for simplicity) are calculated as well: See Figs. 3, 4 and 5. At high cutoff energies, $\langle E \rangle$ saturates at $2.48944... \times 10^9\;eV$ \cite{cosmic}, a value which is over ten times larger than the Hagedorn temperature (close to $1.8 \times 10^8 \;eV$ \cite{beckannihilation}). In the same limit,  we obtain for the specific-heat-like quantity $M_2 \simeq \langle E^2 \rangle \simeq 6.29\times 10^{21} \,(eV)^2$. Finally, $M_3  \simeq \langle E^3 \rangle$ diverges with increasing $E_{cutoff}$. This is of course due to the fact that, in the high energy limit, $\Phi \propto 1/E^{\frac{1}{q-1} -2} \sim 1/E^{3.4}$; consequently the third moment integrand vanishes like $1/E^{0.4}$, which is not integrable at infinity.  

One may guess that, along such wide ranges (of both fluxes and energies), a variety of complex intra- and inter-galactic phenomena are involved, related to both the sources of the cosmic rays as well as the media they cross before arriving on Earth. However, from a phenomenological viewpoint, the overall results amount to something quite simple. Indeed, by solving a simple differential equation, a quite remarkable agreement is obtained \cite{cosmic}. This differential equation is the following one:
\beq
\frac{dp_i}{dE_i}= -b^\prime p_i^{q^\prime} -b p_i^q \;.
\eeq
This differential equation has remarkable particular cases. The most famous one is $(q^\prime,q)=(1,2)$, since it precisely corresponds to the differential equation which enabled Planck, in his October 1900 paper, to (essentially) guess the black-body radiation distribution, thus opening (together with his December 1900 paper) the road to quantum mechanics. The more general case $q^\prime=1$ and arbitrary $q$ is a simple particular instance of the Bernoulli equation, and, as such, has a simple explicit solution (Eq. (20) with $a_1=-b^\prime$ and $a_q=-b$). This solution has proved its efficiency in a variety of problems, including in generalizing the Zipf-Mandelbrot law for quantitative linguistics (for a review, see Montemurro's article in the Gell-Mann--Tsallis volume \cite{tsallis3}). Finally, the generic case $q>q^\prime>1$ also has an explicit solution (though not particularly simple, but in terms of two hypergeometric functions; see \cite{bemski}) and produces, taking also into account the ultra-relativistic ideal gas density of states, the above mentioned quite good agreement with the observed fluxes. Indeed, if we assume $0<b^\prime<<b$ and $q^\prime<q$, the distribution makes a neat crossover from a power-law characterized by $q$ at low energies to a power-law characterized by $q^\prime$ at high energies, which is exactly what the cosmic rays exhibit to a quite good approximation.   Let us finally mention that the first possible microscopic interpretation of our phenomenological theory has just been suggested \cite{beckcosmic}.
   
For possible effects of a slightly nonextensive black-body radiation on cosmic rays see \cite{anchordoqui}. Finally, other aspects related to cosmic rays have been shown to exhibit fingerprints of nonextensivity \cite{wilkcosmic}. 
 
\section{conclusions}

In nuclear and high energy physics, there is a considerable amount of anomalous phenomena that benefit from a thermostatistical treatment which exceeds the usual capabilities of Boltzmann-Gibbs statistical mechanics. This fact is due to the relevance of long-range forces, as well as to a variety of dynamical nonlinear dynamical aspects, possibly leading to nonmarkovian processes, i.e., long-term microscopic memory. Some of these phenomena appear to be tractable within nonextensive statistical mechanics, and we have illustrated this with a few typical examples. For the particular case of cosmic rays, we have indicated their average energy $\langle E \rangle \simeq 2.48944... \times 10^9\;eV$, and the specific-heat-like quantity $\langle E^2 \rangle - \langle E \rangle^2 \simeq 6.29\times 10^{21} \,(eV)^2$, with the hope that they might be usefully compared to related astrophysical quantities, either already available in the literature, or to be studied. Along the same vein we have also presented the dependence of such momenta on a possibly existing high-energy cutoff.

In addition to this, we have sketched the possible generalization of nonextensive statistical mechanics on the basis of a recently introduced entropic form \cite{andreconstantino}, whose stationary state is the Beck-Cohen superstatistics \cite{beckcohen}. The metaequilibrium distribution associated with a crossover between $q$-statistics and $q^\prime$-statistics can be seen as a particular case of this generalized nonextensive statistical mechanics.  

It is worthy to mention at this point that the present attempts for further generalization of BG statistical mechanics are to be understood on physical grounds, and by no means as informational quantities that can be freely introduced to deal with specific tasks, and which can in principle depend on as many free (or fitting) parameters as one wishes. Examples of such informational quantities are the Renyi entropy (depending on one parameter and being usefully applied in the multifractal characterization of chaos), Sharma-Mittal entropy (which contains both Renyi entropy and $S_q$ as particular cases), and very many others that are available in the literature. The precise criteria for an entropic form to be considered a possible physical entropy are yet not fully understood, {\it although it is already clear that it must have a microscopic dynamical foundation}. It seems however reasonable to exclude, at this stage, those forms which, {\it in contrast with $S_{BG}$ and $S_q$}, (i) are {\it not} concave (or convex), since this would seriously damage the capability for thermodynamical stability and for satisfactory thermalization of different systems put in thermodynamical contact, or (ii) are {\it not} stable, since this would imply that the associated quantity would not be robust under experimental measurements. These crucial points, and several others (probably equally important, such as the finite entropy production per unit time), have been disregarded by Luzzi et al \cite{luzzi} in their recent criticism of nonextensive statistical mechanics. Indeed, Renyi entropy, Sharma-Mittal  entropy (that Luzzi et al mention without any justification at the same epistemological level as $S_q$) {\it are neither concave nor stable for arbitrary values of their parameters}. These and other information measures (most of them not concave and/or not stable) have been freely introduced, along various decades, as optimizing tools for specific tasks. They can certainly be useful for various purposes, which do {\it not} necessarily include the specific one we are addressing here: a thermodynamically meaningful generalization of the Boltzmann-Gibbs physical entropy.

\section*{Acknowledgments}
We are indebted to T. Kodama, G. Wilk, I. Bediaga, E.G.D. Cohen, M. Baranger and J. Anjos for useful remarks that we have received along the years. One of us (CT) is grateful to M. Gell-Mann for many and invaluable discussions on this subject. This work has been partially supported by PRONEX/MCT, CNPq, and FAPERJ (Brazilian agencies).

\newpage
\begin{figure}[htb]
\begin{center}
\includegraphics[width=0.75\textwidth,keepaspectratio,clip=]{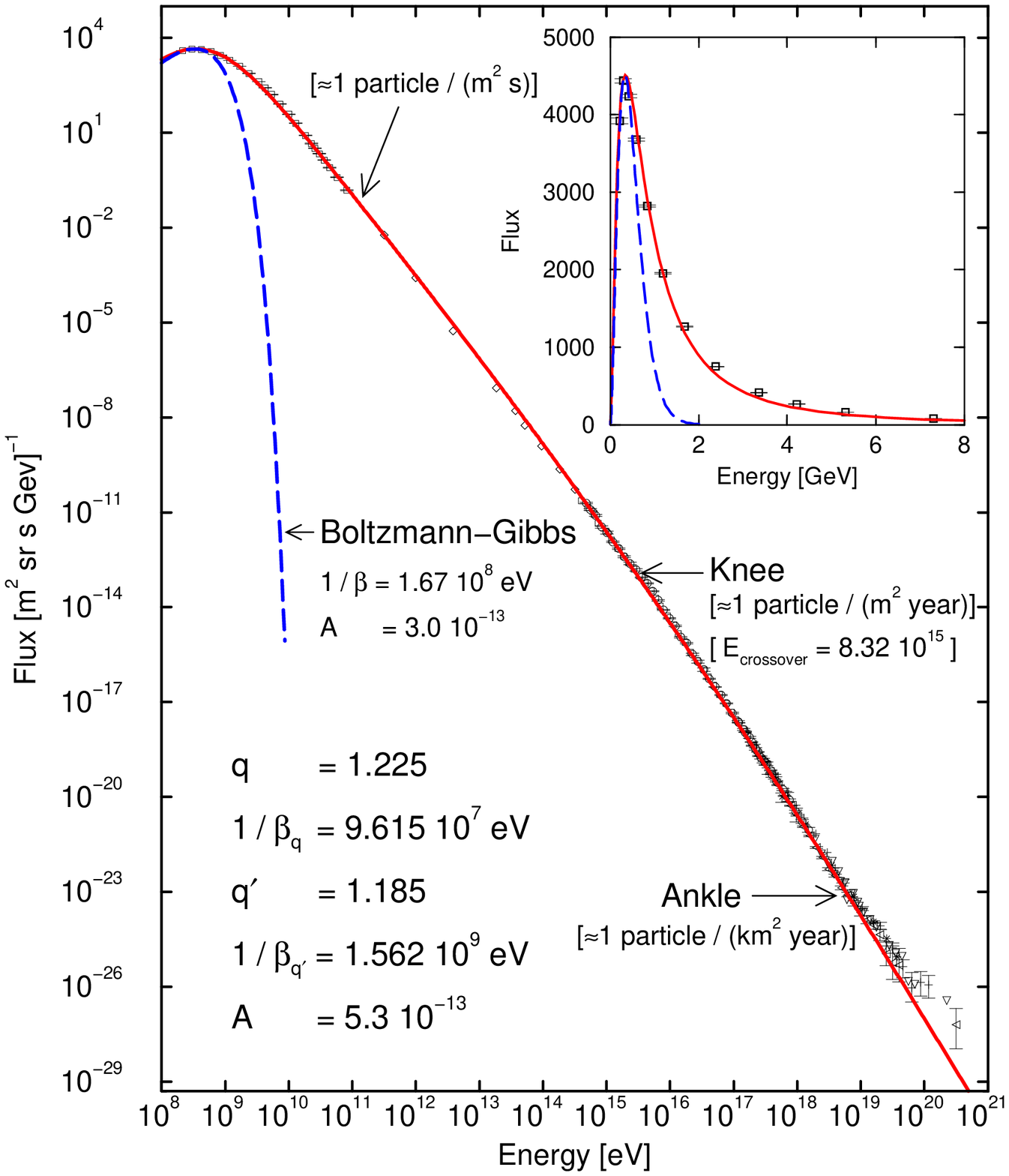}
\end{center}
\caption{\small
Flux of cosmic rays as function of their energy. See [15] for details.}
\label{Fig_1}
\end{figure}

\newpage
\begin{figure}[htb]
\begin{center}
\includegraphics[width=0.75\textwidth,keepaspectratio,clip=]{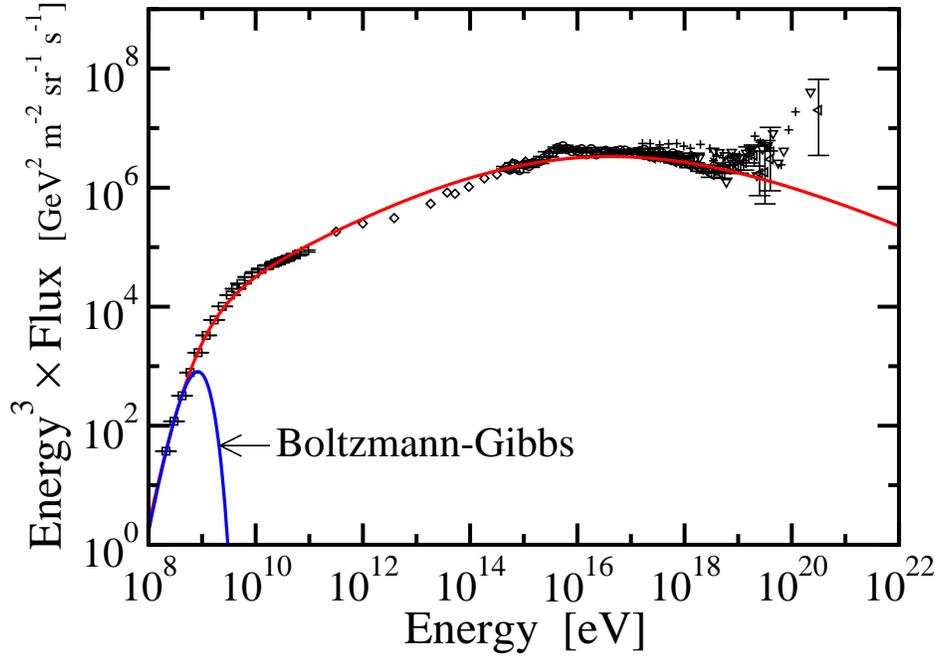}
\end{center}
\caption{\small 
Same as in Fig. 1, but the ordinate is now multiplied by $(Energy)^3$.}
\label{Fig_2}
\end{figure}

\begin{figure}[htb]
\begin{center}
\includegraphics[width=0.75\textwidth,keepaspectratio,clip=]{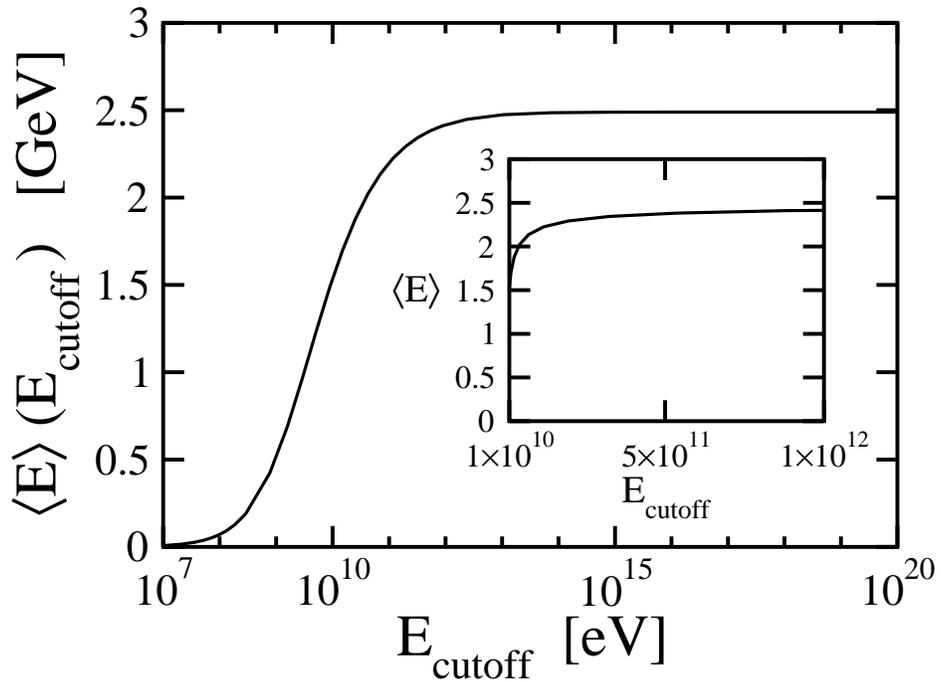}
\end{center}
\caption{\small 
$\langle E \rangle$ as a function of the cutoff energy.}
\label{Fig_3}
\end{figure}

\newpage
\begin{figure}[htb]
\begin{center}
\includegraphics[width=0.75\textwidth,keepaspectratio,clip=]{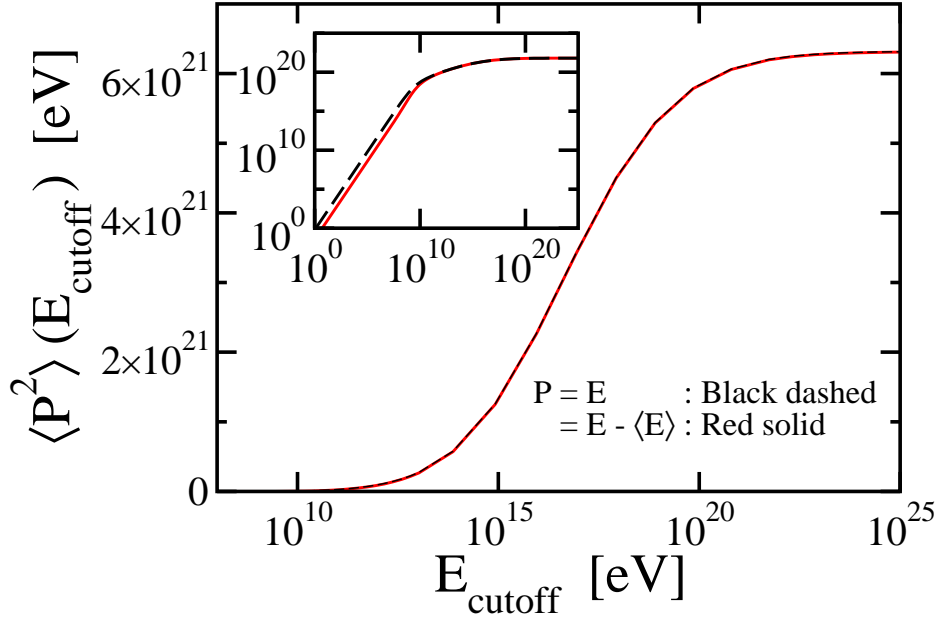}
\end{center}
\caption{\small 
$\langle E^2 \rangle$ (black dashed curves) and $M_2 \equiv \langle E^2 \rangle - \langle E\rangle^2$ (red continuous curves) as functions of the cutoff energy.}
\label{Fig_4}
\end{figure}

\begin{figure}[htb]
\begin{center}
\includegraphics[width=0.75\textwidth,keepaspectratio,clip=]{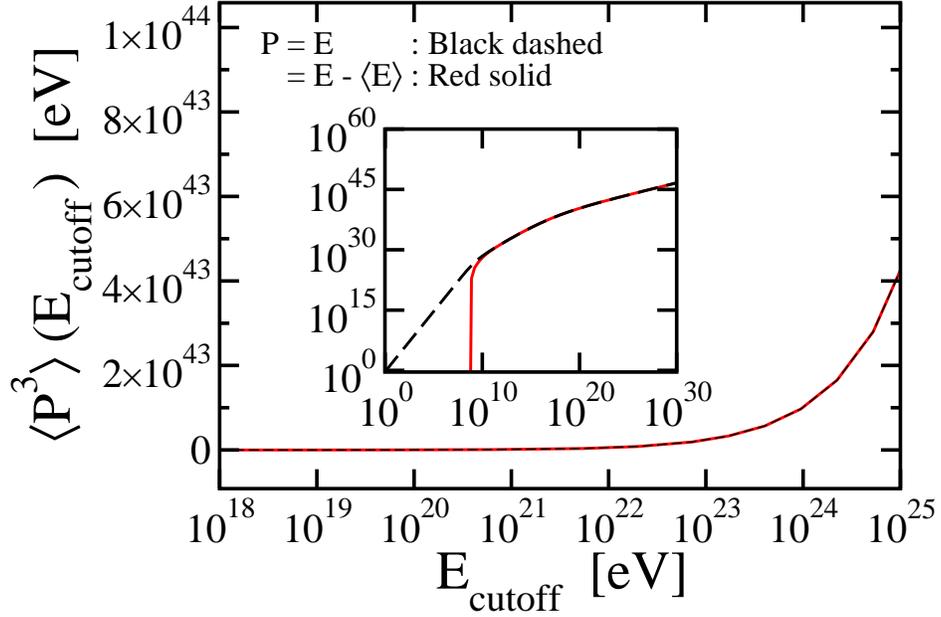}
\end{center}
\caption{\small 
$\langle E^3 \rangle$ (black dashed curves) and $M_3 \equiv \langle E^3 \rangle - 3 \langle E \rangle \langle E\rangle^2 + 2 \langle E \rangle^3 $ (red continuous curves) as functions of the cutoff energy. At vanishing $E_{cutoff}$, $M_3$ vanishes from below, i.e., with slightly negative values.}
\label{Fig_5}
\end{figure}

\begin{thebibliography}{10}

\bibitem{einstein}A. Einstein, Annalen der Physik {\bf 33}, 1275 (1910) [Translation: Abraham Pais, {\it
Subtle is the Lord...}, Oxford University Press, 1982)].

\bibitem{tsallis1}C. Tsallis, J. Stat. Phys. {\bf 52}, 479 (1988).

\bibitem{cohen} E.G.D. Cohen, Physica A {\bf 305}, 19 (2002).

\bibitem{baranger} M. Baranger, Physica A {\bf 305}, 27 (2002).

\bibitem{turbulentbeck}C. Beck, G. S. Lewis and H. L. Swinney,
Phys. Rev. E {\bf 63}, 035303 (2001); C. Beck, Phys. Rev. Lett. {\bf 87}, 180601 (2001); C. Beck, Europhys. Lett. {\bf 57}, 329 (2002).

\bibitem{turbulentarimitsu}T. Arimitsu and N. Arimitsu, Physica A {\bf 305}, 218 (2002).

\bibitem{bediaga}I. Bediaga, E. M. F. Curado and J. Miranda, Physica A {\bf 286}, 156 (2000).

\bibitem{beckannihilation}C. Beck, Physica A {\bf 286}, 164 (2000). See also C. Beck, Physica D {\bf 171}, 72 (2002).

\bibitem{wilk}M. Rybczynski, Z. Wlodarczyk and G. Wilk, {\it Rapidity spectra analysis in terms of non-extensive statistic approach}, preprint (2002) [hep-ph/0206157];  O.V. Utyuzh, G. Wilk and Z. Wlodarczyk, J. Phys. G {\bf 26}, L39 (2000); G. Wilk and Z. Wlodarczyk, Phys. Rev. Lett. {\bf 84}, 2770 (2000);  G. Wilk and Z. Wlodarczyk, in {\it Non Extensive Statistical Mechanics and Physical Applications}, eds. G. Kaniadakis, M. Lissia and A. Rapisarda,  Physica A {\bf 305}, 227 (2002);  G. Wilk and Z. Wlodarczyk,  in {\it Classical and Quantum Complexity and Nonextensive Thermodynamics}, eds. P. Grigolini, C. Tsallis and B.J. West, Chaos, Solitons and Fractals {\bf 13}, Number 3, 547 (Pergamon-Elsevier, Amsterdam, 2002); F.S. Navarra, O.V. Utyuzh, G. Wilk and Z. Wlodarczyk, N. Cimento C {\bf 24}, 725 (2001); G. Wilk and Z. Wlodarczyk, in Proc. 6th International Workshop on Relativistic Aspects of Nuclear Physics (RANP2000, Tabatinga, Sao Paulo, Brazil, 17-20 October 2000); R. Korus, St. Mrowczynski, M. Rybczynski and Z. Wlodarczyk, Phys. Rev. C {\bf 64}, 054908 (2001); 
G. Wilk and Z. Wlodarczyk, {\it Traces of nonextensivity in particle physics due to fluctuations}, ed. N. Antoniou  (World Scientific, 2003), to appear [hep-ph/0210175].

\bibitem{kodama}C.E. Aguiar and T. Kodama, {\it Nonextensive statistics and multiplicity distribution in hadronic collisions}, Physica A (2003), in press.

\bibitem{ion}D.B. Ion and M.L.D. Ion, Phys. Rev. Lett. {\bf 81}, 5714 (1998); M.L.D. Ion and D.B. Ion, Phys. Rev. Lett. {\bf 83}, 463 (1999); D.B. Ion and M.L.D. Ion, Phys. Rev. E {\bf 60}, 5261 (1999); D.B. Ion and M.L.D. Ion, in {\it Classical and Quantum Complexity and Nonextensive Thermodynamics}, eds. P. Grigolini, C. Tsallis and B.J. West, Chaos , Solitons and Fractals {\bf 13}, Number 3, 547 (Pergamon-Elsevier, Amsterdam, 2002); M.L.D. Ion and D.B. Ion, Phys. Lett. B {\bf 474}, 395 (2000); M.L.D. Ion and D.B. Ion, Phys. Lett. B {\bf 482}, 57 (2000); D.B. Ion and M.L.D. Ion, Phys. Lett. B {\bf 503}, 263 (2001).

\bibitem{superkamiokande}M. Coraddu, M. Lissia, G. Mezzorani and P. Quarati, {\it Super-Kamiokande hep neutrino best fit: A possible signal of nonmaxwellian solar plasma}, Physica A (2003), in press [hep-ph/0212054].

\bibitem{neutrino}G. Kaniadakis, A. Lavagno and P. Quarati, Phys. Lett. B {\bf 369}, 308 (1996); P. Quarati, A. Carbone, G. Gervino, G. Kaniadakis, A. Lavagno and E. Miraldi, Nucl. Phys. A {\bf 621}, 345c (1997); G. Kaniadakis, A. Lavagno and P. Quarati, Astrophysics and space science {\bf 258}, 145 (1998); G. Kaniadakis, A. Lavagno, M. Lissia and P. Quarati, in Proc. 5th International Workshop on {\it Relativistic Aspects of Nuclear Physics} (Rio de Janeiro-Brazil, 1997); eds. T. Kodama, C.E. Aguiar, S.B. Duarte, Y. Hama, G. Odyniec and H. Strobele (World Scientific, Singapore, 1998), p. 193;  M. Coraddu, G. Kaniadakis, A. Lavagno, M. Lissia, G. Mezzorani and P. Quarti, in {\it Nonextensive Statistical Mechanics and Thermodynamics}, eds. S.R.A. Salinas and C. Tsallis, Braz. J. Phys. {\bf 29}, 153 (1999); A. Lavagno and P. Quarati, Nucl. Phys. B, Proc. Suppl. {\bf 87}, 209 (2000); C.M. Cossu, {\it Neutrini solari e statistica di Tsallis}, Master Thesis, Universita degli Studi di Cagliari (2000).

\bibitem{rafelski}D.B. Walton and J. Rafelski, Phys. Rev. Lett. {\bf 84}, 31 (2000).

\bibitem{cosmic}C. Tsallis, J.C. Anjos and E.P. Borges, {\it Fluxes of cosmic rays: A delicately balanced anomalous stationary state}, astro-ph/0203258 (2002).

\bibitem{hagedorn}R. Hagedorn, N. Cim. {\bf 3}, 147 (1965).

\bibitem{polytrope}A.R. Plastino and A. Plastino, Phys. Lett. A {\bf 174}, 384 (1993); J.-J. Aly and J. Perez, Phys. Rev. E {\bf 60}, 5185 (1999); A. Taruya and M. Sakagami, Physica A {\bf 307}, 185 (2002); A. Taruya and M. Sakagami, {\it Gravothermal catastrophe and Tsallis' generalized entropy of self-gravitating systems II. Thermodynamic properties of stellar polytrope}, Physica A (2003), in press [cond-mat/0204315]; P.H. Chavanis, {\it Gravitational instability of isothermal and polytropic spheres}, Astronomy and Astrophysics (2003), in press [astro-ph/0207080]; P.-H. Chavanis, Astro. and Astrophys. {\bf 386}, 732 (2002).

\bibitem{peculiar}A. Lavagno, G. Kaniadakis, M. Rego-Monteiro, P. Quarati and C. Tsallis, Astrophysical Letters and Communications {\bf  35}, 449 (1998).

\bibitem{cosmology}
V.H. Hamity and D.E. Barraco, Phys. Rev. Lett. {\bf  76}, 4664 (1996); V.H. Hamity and D.E. Barraco, Physica A {\bf 282}, 203 (2000); L.P. Chimento, J. Math. Phys. {\bf 38}, 2565 (1997); D.F. Torres, H. Vucetich and A. Plastino, Phys. Rev. Lett. {\bf 79}, 1588 (1997) [Erratum: {\bf 80}, 3889 (1998)]; U. Tirnakli and D.F. Torres, Physica A  {\bf 268}, 225 (1999); L.P. Chimento, F. Pennini and A. Plastino, Physica A {\bf 256}, 197 (1998); L.P. Chimento, F. Pennini and A. Plastino, Phys. Lett. A {\bf 257}, 275 (1999); D.F. Torres and H. Vucetich, Physica A {\bf 259}, 397 (1998); D.F. Torres, Physica A {\bf 261}, 512 (1998); H.P. de Oliveira, S.L. Sautu, I.D. Soares and E.V. Tonini, Phys. Rev. D {\bf 60}, 121301-1 (1999); H.P. de Oliveira, I.D. Soares and E.V. Tonini, Physica A {\bf 295}, 348 (2001); M.E. Pessah, D.F. Torres and H. Vucetich, Physica A {\bf 297}, 164 (2001); M.E. Pessah and D.F. Torres, Physica A {\bf 297}, 201 (2001); C. Hanyu and A. Habe, Astrophys. J. {\bf 554}, 1268 (2001); E.V. Tonini, {\it Caos e universalidade em modelos cosmologicos com pontos criticos centro-sela}, Doctor Thesis (Centro Brasileiro de Pesquisas Fisicas, Rio de Janeiro, March 2002)

\bibitem{astero}G.A. Tsakouras, A. Provata and C. Tsallis, {\it Non-extensive properties of the cyclic lattice Lotka-Volterra model}, in preparation (2003).

\bibitem{economics1}C. Anteneodo, C. Tsallis and A.S. Martinez, Europhys. Lett. {\bf 59}, 635 (2002).

\bibitem{economics2}L. Borland, Phys. Rev. Lett. {\bf 89}, 098701 (2002);  Quantitative Finance {\bf 2}, 415 (2002). 

\bibitem{economics3}R. Osorio, L. Borland and C. Tsallis, in {\it Nonextensive Entropy - Interdisciplinary Applications}, M. Gell-Mann and C. Tsallis, eds. (Oxford University Press, 2003), in preparation; see also F. Michael and M.D. Johnson, {\it Financial marked dynamics}, Physica A (2003), in press. 

\bibitem{hydra}A. Upadhyaya, J.-P. Rieu, J.A. Glazier and Y. Sawada, Physica A {\bf 293}, 549 (2001).

\bibitem{ademir}J. A. S. de Lima, R. Silva and A. R. Plastino, Phys. Rev. Lett. {\bf 86}, 2938 (2001).

\bibitem{chaos1}C. Tsallis, A.R. Plastino and W.-M. Zheng, Chaos, Solitons \& Fractals {\bf 8}, 885 (1997); U.M.S. Costa, M.L. Lyra, A.R. Plastino and C. Tsallis,
Phys. Rev. E {\bf 56}, 245 (1997); M.L. Lyra and C. Tsallis, Phys. Rev. Lett. {\bf 80}, 53 (1998); U. Tirnakli, C. Tsallis and M.L. Lyra, Eur. Phys. J. B {\bf 11}, 309 (1999); V. Latora, M. Baranger, A. Rapisarda, C. Tsallis, Phys. Lett. A {\bf 273}, 97 (2000); F.A.B.F. de Moura, U. Tirnakli, M.L. Lyra, Phys. Rev. E {\bf 62}, 6361 (2000); U. Tirnakli, G. F. J. Ananos, C. Tsallis, Phys. Lett. A {\bf 289}, 51 (2001); H. P. de Oliveira, I. D. Soares and E. V. Tonini, Physica A {\bf 295}, 348 (2001); F. Baldovin and A. Robledo, Europhys. Lett. {\bf 60}, 518 (2002);  F. Baldovin and A. Robledo, Phys. Rev. E {\bf 66}, 045104(R) (2002); E.P. Borges, C. Tsallis, G.F.J. Ananos and P.M.C. Oliveira, Phys. Rev. Lett. {\bf 89}, 254103 (2002); U. Tirnakli, Physica A {\bf 305}, 119 (2002); U. Tirnakli, Phys. Rev. E {\bf 66}, 066212 (2002). 

\bibitem{chaos2}Y. Weinstein, S. Lloyd and C. Tsallis, Phys. Rev. Lett. {\bf 89}, 214101 (2002).

\bibitem{entanglement}S. Abe and A.K. Rajagopal, Physica A {\bf 289}, 157 (2001), C. Tsallis; S. Lloyd and M. Baranger, Phys. Rev. A {\bf 63}, 042104 (2001); C. Tsallis, P.W. Lamberti and D. Prato, Physica A {\bf 295}, 158 (2001); F.C. Alcaraz and C. Tsallis, Phys. Lett. A {\bf 301}, 105 (2002); C. Tsallis, D. Prato and C. Anteneodo, Eur. Phys. J. B {\bf 29}, 605  (2002); J. Batle, A.R. Plastino, M. Casas and A. Plastino, {\it Conditional $q$-entropies and quantum separability: A numerical exploration}, quant-ph/0207129 (2002).

\bibitem{nonlinearFP}A.R. Plastino and A. Plastino, Physica A  {\bf 222}, 347 (1995); C. Tsallis and D.J. Bukman, Phys. Rev. E {\bf 54}, R2197 (1996); C. Giordano, A.R. Plastino, M. Casas and A. Plastino, Eur. Phys. J. B {\bf 22}, 361 (2001); A. Compte and D. Jou, J. Phys. A {\bf  29}, 4321 (1996); A.R. Plastino, M. Casas and A. Plastino, Physica A {\bf 280}, 289 (2000); M. Bologna, C. Tsallis and P. Grigolini, Phys. Rev. E {\bf 62}, 2213 (2000); C. Tsallis and E.K. Lenzi, in {\it Strange Kinetics}, eds. R. Hilfer et al, Chem. Phys.  {\bf 284}, 341 (2002) [Erratum (2002)]; E.K. Lenzi, L.C. Malacarne, R.S. Mendes and I.T. Pedron, {\it Anomalous diffusion, nonlinear fractional Fokker-Planck equation and solutions}, cond-mat/0208332 (2002); E.K. Lenzi, C. Anteneodo and L. Borland, Phys. Rev. E {\bf 63}, 051109 (2001); E.M.F. Curado and F.D. Nobre, {\it Derivation of nolinear Fokker-Planck equations by means of approximations to the master equation}, Phys. Rev. E {\bf 67}, 0211XX (2003), in press; C. Anteneodo and C. Tsallis, {\it Multiplicative noise: A mechanism leading to nonextensive statistical mechanics}, cond-mat/0205314 (2002).

\bibitem{longrange}C. Anteneodo and C. Tsallis, Phys. Rev. Lett {\bf 80}, 5313 (1998); V. Latora, A. Rapisarda and C. Tsallis, Phys. Rev. E {\bf 64}, 056134 (2001); 
A. Campa, A. Giansanti and D. Moroni, in {\it Non Extensive Statistical Mechanics and Physical Applications}, eds. G. Kaniadakis, M. Lissia and A. Rapisarda,  Physica A {\bf 305}, 137 (2002); B.J.C. Cabral and C. Tsallis, Phys. Rev. E {\bf 66}, 065101(R) (2002); E.P. Borges and C. Tsallis, in {\it Non Extensive Statistical Mechanics and Physical Applications}, eds. G. Kaniadakis, M. Lissia and A. Rapisarda,  Physica A {\bf 305}, 148 (2002); A. Campa, A. Giansanti, D. Moroni and C. Tsallis, Phys. Lett. A {\bf 286}, 251 (2001); M.-C. Firpo and S. Ruffo, J. Phys. A {\bf 34}, L511 (2001); C. Anteneodo and R.O. Vallejos, Phys. Rev. E  {\bf 65}, 016210 (2002); R.O. Vallejos and C. Anteneodo, Phys. Rev. E {\bf 66}, 021110 (2002); M.A. Montemurro, F. Tamarit and C. Anteneodo, {\it Aging in an infinite-range Hamiltonian of coupled rotators}, Phys. Rev. E (2003), in press cond-mat/0205355 (2002).

\bibitem{internet}S. Abe and N. Suzuki,  Phys. Rev. E {\bf 67}, 016106 (2003).

\bibitem{tsallis2}E.M.F. Curado and C. Tsallis, J. Phys. A: Math. Gen. {\bf 24}, L69 (1991) [Corrigenda: 
{\bf 24}, 3187 (1991) and {\bf 25}, 1019 (1992)]; C. Tsallis, R.S. Mendes and A.R. Plastino, Physica A {\bf 261}, 534 (1998).

\bibitem{tsallis3}S.R.A. Salinas and C. Tsallis, eds., {\it Nonextensive Statistical Mechanics and Thermodynamics}, Braz. J. Phys. {\bf 29}, No. 1 (1999);
S. Abe and Y. Okamoto, eds., {\it Nonextensive Statistical Mechanics and its Applications}, Series {\it Lecture Notes in Physics} (Springer-Verlag, Berlin, 2001); G. Kaniadakis, M. Lissia and A. Rapisarda, eds., {\it Non Extensive Statistical Mechanics and Physical Applications}, Physica A {\bf 305}, No 1/2 (Elsevier, Amsterdam, 2002); M. Gell-Mann and C. Tsallis, eds., {\it Nonextensive Entropy - Interdisciplinary Applications} (Oxford University Press, 2003), in preparation; H.L. Swinney and C. Tsallis, eds.,  {\it Anomalous Distributions, Nonlinear Dynamics, and Nonextensivity}, Physica D (2003), in preparation. An updated bibliography can be found at the web site
http://tsallis.cat.cbpf.br/biblio.htm

\bibitem{beckcohen}C. Beck and E.G.D. Cohen,  {\it Superstatistics}, Physica A (2003), in press [cond-mat/0205097].

\bibitem{abestable}S. Abe, Phys. Rev. E {\bf 66}, 046134 (2002).

\bibitem{andreconstantino}C. Tsallis and A.M.C. Souza, {\it Constructing a statistical mechanics for Beck-Cohen superstatistics}, Phys. Rev. E {\bf 67}, 0261XX (1 Feb 2003), in press  [cond-mat/0206044].

\bibitem{stablesuperstatistics}A.M.C. Souza and C. Tsallis, {\it Stability of the entropy for superstatistics}, preprint (2003) [cond-mat/0301304].

\bibitem{bemski}C. Tsallis, G. Bemski and  R.S. Mendes, Phys. Lett. {\bf 257}, 93 (1999). 

\bibitem{beckcosmic}C. Beck, {\it Generalized statistical mechanics of cosmic rays}, preprint (2003) [cond-mat/0301354].

\bibitem{anchordoqui}L.A. Anchordoqui and D.F. Torres, Phys. Lett. A {\bf 283}, 319 (2001).

\bibitem{wilkcosmic}G. Wilk and Z. Wlodarcsyk, Nucl. Phys. B  (Proc. Suppl.) {\bf 75A}, 191 (1999); G. Wilk and Z. Wlodarczyk, {\it Nonexponential decays and nonextensivity}, Phys. Lett. A {\bf 290}, 55 (2001).

\bibitem{luzzi}R. Luzzi, A.R. Vasconcellos and J.G. Ramos, {\it On the question of the so-called ``Nonextensive thermo-statistics"}, preprint (2002, IFGW-Unicamp internal report); Science {\bf 298}, 1171 (2002).


\end{thebibliography}
\end{document}